# One-dimensional Hubbard-Luttinger model for carbon nanotubes

## H.A. Ishkhanyan[1,2] and V.P. Krainov[1]

[1]Moscow Institute of Physics and Technology, 141700 Dolgoprudny, Russia
[2]Institute for Physical Research, NAS of Armenia, 0203 Ashtarak, Armenia

**Abstract.** A Hubbard-Luttinger model is developed for qualitative description of one-dimensional motion of interacting Pi-conductivity-electrons in carbon single-wall nanotubes at low temperatures. The low-lying excitations in one-dimensional electron gas are described in terms of interacting bosons. The Bogolyubov transformation allows one to describe the system as an ensemble of non-interacting quasi-bosons. Operators of Fermi-excitations and Green functions of fermions are introduced. The electric current is derived as a function of potential difference on the contact between a nanotube and a normal metal. Deviations from Ohm law produced by electron-electron short-range repulsion as well as by the transverse quantization in single-wall nanotubes are discussed. The results are compared with experimental data.

**PACS numbers**: 73.23.Ad Ballistic transport, 72.15.Nj Collective modes (e.g., in one-dimensional conductors), 73.40.Cg Contact resistance, contact potential, 73.63.Fg Nanotubes

**Keywords:** Hubbard model, Luttinger liquid, carbon single-wall nanotubes, electrical conductivity

## 1. Introduction

We consider a system of interacting electrons in one-dimensional approximation. It is known that the standard Landau Fermi-liquid theory of interacting fermions is inapplicable to the one-dimensional case. In this case the one-dimensional Hubbard model [1,2] is applied, which relies on the following two basic assumptions: (i) strong repulsion between two electrons in the same narrow potential well and (ii) small probability for the electron jump to the neighboring well. Another known model, the Luttinger-liquid model [3,4], allows one to analytically describe the one-dimensional system of electrons with short-range repulsion between them at low temperature using two other assumptions: (i) all electrons have energies near the Fermi level $\varepsilon_F$, therefore the energy spectrum is linear: $\varepsilon = (p - p_F) p_F$ (here and hereafter the electron mass is assumed $m = 1$), and (ii) after the collisions with each other the electrons may move either in the same direction (transferred momentum then is $\Delta p = 0$) or in the opposite direction (transferred momentum in this case is $\Delta p = 2 p_F$).

In the Luttinger model, even weak Coulomb interactions cause strong perturbations. For instance, tunneling into a Luttinger liquid at energies near the Fermi level is predicted to be strongly suppressed, unlike what happens in the case of two- and three-dimensional metals. Besides, the differential conductivity scales as power law with respect to bias voltage [5]. Thus, one may expect that the electrically conducting single-wall carbon nanotubes may exhibit Luttinger-liquid behavior.

On the other hand, one should account for the bound electrons' influence on the pure Luttinger behavior. To discuss the role of this subsystem of electrons, one may apply the one-dimensional Hubbard model [2]. Usually, such a treatment uses the Bethe ansatz, which suggests a convenient variational wave function for a many-particle system [6]. Using this approach, a gas of one-dimensional Bose-particles interacting via a repulsive delta-function potential has been considered in [7]. The energies and wave functions for the ground state and low-lying excited states of a system of one-dimensional fermions also interacting via a repulsive delta function potential have been calculated in [8].

In the present paper, a Hubbard-Luttinger model is developed for qualitative description of one-dimensional motion of interacting Pi-conductivity-electrons in carbon single-wall nanotubes at low temperatures. The low-lying excitations in one-dimensional electron gas are described in terms of interacting bosons. Using the Bogolyubov transformation, the system is further described as an ensemble of non-interacting quasi-bosons. Then operators of Fermi-excitations and Green functions of fermions are introduced. Finally, the electric current is calculated as a function of voltage on contact between a nanotube and a normal metal. Deviations from Ohm law produced by electron-electron short-range repulsion [9] as well as by the transverse quantization in single-wall nanotubes [10] are discussed. Comparison of the obtained results with experimental data of [11] shows qualitative agreement in quantum interference oscillations of conductivity.

## 2. Simplification of the Hubbard model

In order to simplify the Hubbard model, we first consider the two-electron Schrödinger equation with the Hamiltonian (here and hereafter we put $\hbar = m = k = 1$)

$$\hat{H}(x_1, x_2) = -\frac{1}{2}\left(\frac{\partial^2}{\partial x_1^2} + \frac{\partial^2}{\partial x_2^2}\right) + V\delta(x_1 - x_2) - k[\delta(x_1) + \delta(x_2)]. \qquad (1)$$

This Hamiltonian corresponds to two electrons in delta-function potential well with (dimensionless) repulsion potential $V$. The problem is not solved analytically; therefore, we use the variational approach. The symmetrical variational wave function of two electrons with a total spin $S = 0$ can be chosen in the form

$$\Psi(x_1, x_2) = A\left(e^{-\alpha|x_1|-\beta|x_2|} + e^{-\alpha|x_2|-\beta|x_1|}\right). \qquad (2)$$

If $V=0$, one obtains $\alpha=\beta=1$ and the total energy is $E=-1$. The system under consideration is analogous to the negative hydrogen ion. The simplification here is due to the delta-function repulsion between two electrons instead of the Coulomb repulsion.

The result of numerical simulations is that when $V=3$, one of the electrons goes to continuum, while the second electron practically returns to its initial state. In Fig. 1 the energy of two electrons as a function of repulsion potential $V$ is presented. It is seen that the energy increases monotonically with $V$. In Fig. 2 the inverse radius $\alpha$ of the outer electron is shown. It is seen that $\alpha=0$ when $V=3$. The inverse radius $\beta$ of the inner electron is shown in Fig. 3. It is seen that $\beta=1$ when $V=3$. Thus, in this model the (dimensionless) critical repulsion potential is $V=3$. The existence of a critical potential is a known peculiarity of the Hubbard model.

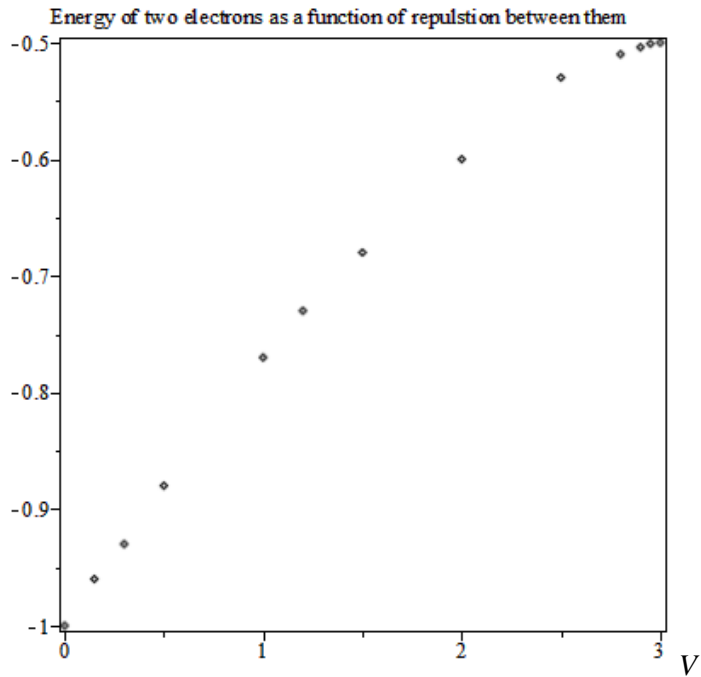

Fig. 1. Dependence of the energy of two electrons on the repulsive potential $V$.

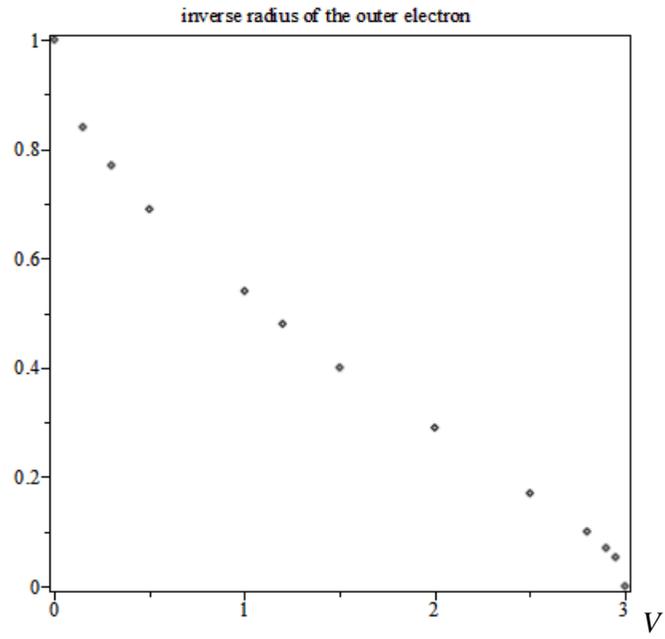

Fig. 2. Dependence of the inverse radius $\alpha$ of the outer electron on the repulsive potential $V$.

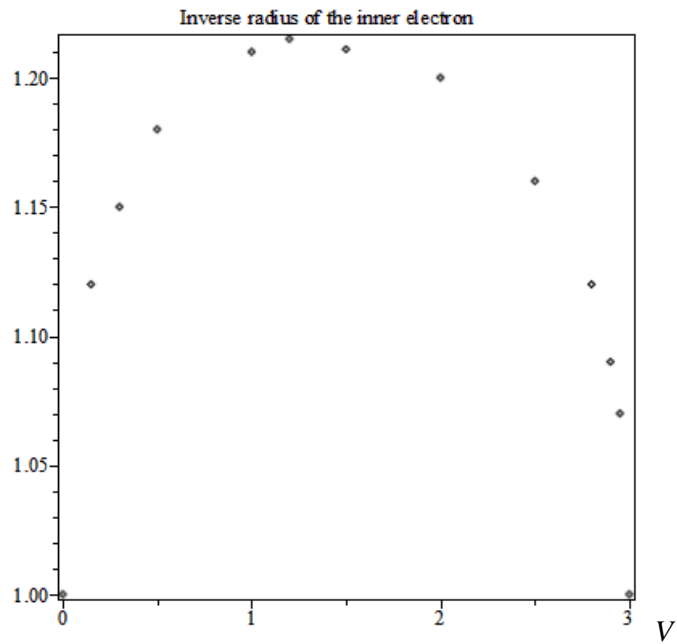

Fig. 3. Dependence of the inverse radius $\beta$ of the inner electron on the repulsive potential $V$.

Suppose further that we have four electrons in two delta-function one-dimensional potential wells with repulsion potential $V > 3$. Based on the above consideration, we may conclude that two electrons should go to the continuum. According to the Pauli principle they should have opposite spins, since their spatial wave functions overlap strongly. Respectively, other two electrons remain in the neighboring potential wells. They also have opposite spins, but, because of the Hubbard assumption, the probability for the electron to jump to the neighboring well is negligible, so that one may disregard the bound electrons. Obviously, this is valid also for the case of a chain of potential wells. As a result, the Hubbard problem reduces to the Luttinger one-dimensional problem of interacting electrons.

### 3. The Luttinger approach

The Luttinger Hamiltonian of interacting electrons is of the form [9]

$$\hat{H} = \hat{H}_0 + \hat{H}_1 + \hat{H}_2,$$

$$\hat{H}_0 = p_F \sum_k (|p_0 + k| - p_0)(\hat{a}^+_{p_0+k}\hat{a}_{p_0+k} + \hat{a}^+_{-p_0-k}\hat{a}_{-p_0-k}),$$

$$\hat{H}_1 = \frac{g_1}{2} \sum_{k_1,k_2,q} \begin{pmatrix} \hat{a}^+_{p_0+k_1+q/2}\hat{a}_{p_0+k_1-q/2}\hat{a}^+_{p_0+k_2-q/2}\hat{a}_{p_0+k_2+q/2} + \\ \hat{a}^+_{-p_0+k_1+q/2}\hat{a}_{-p_0+k_1-q/2}\hat{a}^+_{-p_0+k_2-q/2}\hat{a}_{-p_0+k_2+q/2} \end{pmatrix},$$ (3)

$$\hat{H}_2 = g_2 \sum_{k_1,k_2,q} \hat{a}^+_{p_0+k_1+q/2}\hat{a}_{p_0+k_1-q/2}\hat{a}^+_{-p_0+k_2-q/2}\hat{a}_{-p_0+k_2+q/2}.$$

Here $\hat{H}_0$ describes the kinetic energy of electrons. The term $\hat{H}_1$ describes the scattering of electrons at the collisions with small transferred momentum $q << p_F$, and the term $\hat{H}_2$ stands for the scattering of electrons at the collisions with large transferred momentum $q \approx 2p_F$. In the general case of arbitrary potential, there are two interaction constants in Eq. (3), but in the case of a delta-function potential these constants are equal:

$$g_1 = V_{q=0} = V_{q=2p_F} = g_2 = g, \quad V_q = \int V\delta(x)\exp(-iqx)dx.$$ (4)

So called "right" and "left" density operators corresponding to electron motion to the right or to the left, respectively, are defined in the Luttinger model as:

$$\hat{\rho}_1(q) = \sum_{p>0} \hat{a}^+_{p-q/2}\hat{a}_{p+q/2}, \quad \hat{\rho}_2(q) = \sum_{p<0} \hat{a}^+_{p-q/2}\hat{a}_{p+q/2}.$$ (5)

The Hamiltonian of interacting electrons is then expressed via these operators as

$$\hat{H} = \left(\pi p_F + \frac{g}{2}\right)\sum_q (\hat{\rho}_1(q)\hat{\rho}_1(-q) + \hat{\rho}_2(q)\hat{\rho}_2(-q)) + g\sum_q \hat{\rho}_1(q)\hat{\rho}_2(-q).$$ (6)

If the density operators are expanded into Fourier series:

$$\hat{\rho}_1(q) = \sqrt{\frac{|q|}{2\pi}} \begin{cases} \hat{b}_q; & q > 0 \\ \hat{b}^+_{-q}; & q < 0 \end{cases}, \quad \hat{\rho}_2(q) = \sqrt{\frac{|q|}{2\pi}} \begin{cases} \hat{b}_q; & q < 0 \\ \hat{b}^+_{-q}; & q > 0 \end{cases}, \quad (7)$$

where $\hat{b}_q, \hat{b}^+_{-q}$ are Bose operators, then the Luttinger Hamiltonian takes the form

$$\hat{H} = \left(p_F + \frac{g}{2\pi}\right)\sum_{q>0}\left(\hat{b}^+_q \hat{b}_q + \hat{b}^+_{-q}\hat{b}_{-q}\right) + \frac{g}{2\pi}\sum_{q>0}\left(\hat{b}^+_q \hat{b}^+_{-q} + \hat{b}_q \hat{b}_{-q}\right). \quad (8)$$

The Bogolybov transformation allows one to reduce the problem to an effective one for a system of non-interacting sound bosons:

$$\hat{H} = \sum_q \omega_q \hat{c}^+_q \hat{c}_q, \quad \omega_q = u|q|, \quad u = \sqrt{p_F^2 + \frac{1}{\pi}p_F g}, \quad (9)$$

where the quantity $u$ is the speed of sound for this Hamiltonian.

The right density operators in spatial-time representation are expressed via the Bose operators as

$$\hat{\rho}_1(x,t) = \sum_{q>0}\sqrt{\frac{q}{2\pi}}\{\hat{c}_q \cosh\theta \exp(iq(x-ut)) - \hat{c}^+_{-q}\sinh\theta \exp(iq(x-ut))\} +$$
$$\sum_{q>0}\sqrt{\frac{q}{2\pi}}\{\hat{c}^+_q \cosh\theta \exp(-iq(x-ut)) - \hat{c}_{-q}\sinh\theta \exp(-iq(x-ut))\}, \quad (10)$$

where we have introduced the notation

$$\tanh(2\theta) = g/(g + 2\pi p_F). \quad (11)$$

Similarly, we can express the left density operators in spatial-time representation via the Bose operators.

The next step is the determination of the Fermi operators via right and left density operators:

$$\hat{\psi}_{1,2}(x,t) = (2\pi a)^{-1/2} e^{i\hat{\varphi}_{1,2}(x,t)}, \quad \hat{\varphi}_{1,2}(x,t) = 2\pi \int_{-\infty}^{x}\hat{\rho}_{1,2}(x',t)dx', \quad (12)$$

where $a$ is a small parameter determining the relaxation of the system. It is introduced in order to avoid the divergence of the involved integrals [9]. The right and left Green functions are introduced as

$$G_1(x,t) = \langle\hat{\psi}_1(x,t)\hat{\psi}^+_1(0,0)\rangle = \frac{1}{2\pi a}\langle\exp(i\hat{\varphi}_1(x,t))\exp(-i\hat{\varphi}_1(0,0))\rangle,$$
$$G_2(x,t) = \langle\hat{\psi}_2(x,t)\hat{\psi}^+_2(0,0)\rangle = \frac{1}{2\pi a}\langle\exp(i\hat{\varphi}_2(x,t))\exp(-i\hat{\varphi}_2(0,0))\rangle. \quad (13)$$

The explicit form of these functions is derived by substitution of Eq. (12) into Eq. (13) [9]:

$$G_1(x,t) = \frac{1}{2\pi a}\left(\frac{a}{a-i(x-ut)}\right)^{\cosh^2\theta}\left(\frac{a}{a-i(x+ut)}\right)^{\sinh^2\theta}, \quad (14)$$

$$G_2(x,t) = \frac{1}{2\pi a}\left(\frac{a}{a+i(x+ut)}\right)^{\cosh^2\theta}\left(\frac{a}{a+i(x-ut)}\right)^{\sinh^2\theta}. \quad (15)$$

## 4. The differential tunnel conductivity

The electric current between two one-dimensional Luttinger systems $A$ and $B$ is

$$I(V) = w\int_{-\infty}^{\infty} G_B(x=0,t)G_A(x=0,t)\exp(iVt)dt', \quad (16)$$

where $w$ is the tunneling rate through the contact between nanotubes $A$ and $B$; $V$ is the electric voltage; $x=0$ is the position of the contact. The differential tunnel conductivity $\rho(V) = dI(V)/dV$ is determined by substitution of Eq. (15) into Eq. (16):

$$\rho(V) \sim V^\alpha, \quad \alpha = 2\sinh^2\theta_A + 2\sinh^2\theta_B, \quad \tanh(2\theta_{A,B}) = \frac{g_{A,B}}{g_{A,B}+2\pi p_F} \quad (17)$$

Now we generalize this result by taking into account the transverse quantization in single-wall nanotubes. The Fermi energy is shifted by the quantity [10]

$$E_F \to E_F - \frac{U^2 mR^2}{2\hbar^2} + \frac{\hbar^2(n^2-1/4)}{2mR^2}; \quad n = 0,1,2..., \quad (18)$$

where $R$ is the radius of the nanotube, $UR\delta(r-R)$ is the potential of the well, and the transverse quantization is determined by the integer $n$. Accordingly, the differential tunnel conductivity is modified as

$$\rho(V) = V^\alpha \sum_{n=0}^{\infty} \frac{1}{\sqrt{E_F - \frac{U^2 mR^2}{2\hbar^2} + \frac{\hbar^2(n^2-1/4)}{2mR^2} - V}}. \quad (19)$$

For the typical example of $\pi$-electrons in a single-wall carbon nanotube the involved parameters are as follows:

$$E_F - \frac{U^2 mR^2}{2\hbar^2} = 1\,eV, \quad \alpha = 0.5, \quad R = 0.3\,\text{nm}.$$

With these values, the differential conductivity as a function of the voltage $V$ is shown in Fig. 4. It is seen that the conductivity undergoes pronounced oscillations.

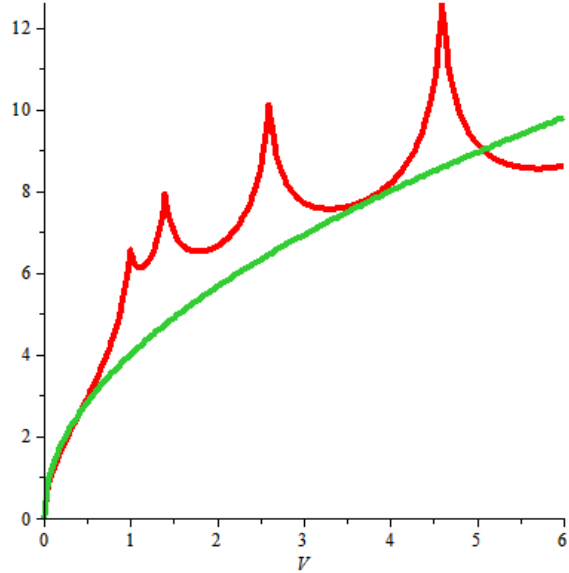

Fig. 4. The conductivity $\rho$ as a function of voltage $V$ (arbitrary units) given by Eq. (19) (red curve). The green curve presents the result without transverse quantization, Eq. (17).

## 5. Conclusion

Thus, we conclude that transverse quantization produces non-monotonic dependence of nanotube conductivity on the voltage in comparison with the standard Luttinger one-dimensional model. The next improvement of the model can be done based on the one-dimensional extended Hubbard model with a weak repulsive short-range interaction in the non-half-filled band case [12]. This approach uses non-perturbative renormalization group methods and Ward identities coming from the asymptotic gauge invariance of the model. At zero temperature the response functions have anomalous power-law decay with logarithmic corrections. A model shows the phenomenon of spin-charge separation, a manifestation of which is that the 2-point function is factorized into the product of two functions. Note that spin-charge separation occurs in the Hubbard model, but is valid only at large distances and up to logarithmic corrections.

The electrical transport properties of well-contacted ballistic single-wall carbon nanotubes at low temperatures have been experimentally studied in [11]. Signatures of strong electron-electron interactions have been observed (the conductivity exhibits bias-voltage-dependent amplitudes of quantum interference oscillations), and the current noise manifests bias-voltage-dependent power-law scalings as was predicted in [9] (see Eq. (17)). We note that Fig. 3 of Ref. [11] demonstrates oscillations in agreement with our predictions given by Eq. (19) and shown in Fig. 4.

**Acknowledgements**

This work was supported by the Russian Foundation for Basic Research (project No. 13-02-00072) and by the State Committee of Science of RA (project No. 13YR-1C0055).